\RequirePackage{silence}
\documentclass[preprint,12pt]{elsarticle}

\usepackage{amssymb}
\usepackage{amsmath}

\usepackage{amsmath}
\usepackage[version=4]{mhchem}
\usepackage[utf8]{inputenc} 
\usepackage{booktabs}
\usepackage{blindtext}

\usepackage[table]{xcolor}

\usepackage{graphicx}
\graphicspath{./images/}

\usepackage{natbib}

\usepackage{color}   
\usepackage{hyperref}
\hypersetup{
    colorlinks=true, 
    linktoc=all,     
    linkcolor=blue,  
}
\usepackage{comment}

\journal{Physical Review E}

\makeatletter
\newcounter{savesection}
\newcounter{apdxsection}
\renewcommand\appendix{\par
  \setcounter{savesection}{\value{section}}%
  \setcounter{section}{\value{apdxsection}}%
  \setcounter{subsection}{0}%
  \gdef\thesection{\@Alph\c@section}}
\newcommand\unappendix{\par
  \setcounter{apdxsection}{\value{section}}%
  \setcounter{section}{\value{savesection}}%
  \setcounter{subsection}{0}%
  \gdef\thesection{\@arabic\c@section}}
\makeatother

\begin{document}
\begin{frontmatter}



\title{What is Adaptable Life?} 


\author{Roger D. Jones $^{a,b,c,d}$}

\affiliation[ECLT]{European Centre for Living Technology (ECLT) Ca' Bottacin, 3911 Dorsoduro Calle Crosera, 30123 Venezia, Italy}
\affiliation[UNIVE]{Dipartimento di Scienze Molecolari e Nanosistemi, Universit\`a Ca' Foscari Venezia, 30123 Venezia, Italy}
\affiliation[UNC]
{Department of Biology, University of North Carolina at Chapel Hill, Chapel Hill, North Carolina 27514, USA}



\affiliation[email]
{Corresponding author: RogerDJonesPhD@gmail.com}

    \begin{abstract}
Since Schr{\"o}dinger's \emph{What Is Life?}, the physical basis of biological organization has been understood in terms of the interplay between matter, energy, and information. Subsequent developments in molecular biology, information theory, nonequilibrium thermodynamics, and evolutionary theory have clarified how hereditary information is stored, maintained, and modified through natural selection. Here, we extend this program by asking what minimal physical principles are required for adaptable life.

We propose six postulates governing adaptive living systems: the existence of an entropy source, longevity of information, fast response to environmental change, repeatable operation, energetic efficiency, and networks of multiple interacting switches. These principles are introduced as a minimal foundation for biological information processing and adaptation. We examine their implications and compare them with observations across multiple levels of biological organization, including genetic inheritance, epigenetic regulation, cellular signaling, neural computation, metabolic networks, and ecological systems.

The resulting framework suggests that adaptability emerges from the interplay of energy flow, information storage, information processing, and natural selection in systems maintained far from thermodynamic equilibrium. Although the proposed principles are qualitative and not yet predictive, they provide a unified perspective on the physical constraints governing adaptive behavior and offer a starting point for the development of a quantitative theory of adaptable life.
\end{abstract}

\end{frontmatter}




\section{What is Life?}
In 1943, Schr\"odinger delivered a series of lectures at Trinity College Dublin that were later published as \emph{What Is Life?} \cite{schrodinger1944}. His central question was how living organisms maintain their highly ordered structure while obeying the laws of thermodynamics. Schr\"odinger argued that life persists by extracting free energy from the environment and exporting entropy, thereby maintaining order in an open system far from thermodynamic equilibrium. Although his notion of ``negative entropy'' is no longer used in a strict thermodynamic sense, the underlying idea anticipated modern views of living systems as nonequilibrium, information-processing structures sustained by continuous energy dissipation \cite{schrodinger1944,goldenfeld2011}.

A second major insight concerned heredity. At a time when the molecular basis of inheritance was unknown, Schr{\"o}dinger proposed that genetic information must be stored in a stable molecular structure he called an ``aperiodic crystal,'' capable of encoding information through the arrangement of its atoms while remaining resistant to thermal fluctuations \cite{schrodinger1944}. This prediction proved remarkably prescient following the discovery of the DNA double helix and its information-carrying properties \cite{watson1953,crick1953}. 

Although many biological details were absent from his analysis, the lectures profoundly influenced the founders of molecular biology and helped establish the view that life should be understood through the interplay of matter, energy, and information. Modern perspectives continue to regard \emph{What Is Life?} as a foundational work linking biology to statistical physics, information theory, and nonequilibrium thermodynamics \cite{goldenfeld2011,ball2011}.

Schr\"odinger's \emph{What Is Life?} explained how hereditary information can be stably stored and maintained in living systems, but it largely omitted the role of natural selection. Several authors have since argued that evolution provides the missing ingredient by explaining how biological information accumulates and adapts over time. In particular, Goldenfeld and Woese recast life as a collective nonequilibrium phenomenon in which heredity, metabolism, and evolution emerge from the dynamics of information-bearing systems driven far from equilibrium \cite{goldenfeld2011}. In this view, natural selection is not separate from physics but is itself a physical process that shapes and preserves information.

A complementary development has been the reformulation of evolutionary theory in information-theoretic terms. Building on the Price equation and related approaches, Frank showed that natural selection can be interpreted as a process by which populations acquire information about their environment through differential reproduction \cite{frank2012}. This perspective connects evolution directly to statistical inference and learning, extending Schr\"odinger's emphasis on information from the storage of hereditary patterns to their adaptive modification through selection.

More recently, nonequilibrium statistical physics has sought to link adaptation and thermodynamics. England proposed that driven systems can become statistically biased toward states that dissipate environmental work more effectively, suggesting a physical basis for the emergence of adaptive structure \cite{england2013}. Together, these developments extend Schr\"odinger's original program by combining heredity, information, nonequilibrium thermodynamics, and natural selection into a unified physical description of living systems.

In this paper, we continue the program initiated by Schr\"odinger by asking what is the minimal intellectual foundation required for adaptable life. We postulate six principles that capture the essential features of adaptive living systems and explore the consequences that follow from them. The postulates are introduced without derivation and are judged solely by the explanatory power of the resulting theory.

From these assumptions, we develop a framework for understanding adaptation, information processing, and nonequilibrium behavior in living systems. We then compare the predicted consequences with biological observations to assess whether the proposed foundation captures essential features of life. The objective is not completeness, but rather to identify a small set of principles from which a broad range of biological phenomena can be understood.

\section{What is Natural Selection?}
From a physicist's perspective, natural selection is a statistical process that causes information about the environment to accumulate in a population of replicating systems. Whenever replication, variation, and differential reproduction are present, the probability distribution over population states changes in time. Variants that reproduce more successfully become increasingly probable, while less successful variants are suppressed. In this sense, natural selection can be viewed as a dynamical process that biases populations toward regions of state space that are better matched to their environment \cite{frank2012}.

An information-theoretic interpretation makes this picture more precise. The environment continually tests competing variants, and successful variants leave more descendants, causing information about environmental conditions to become encoded in the population. Evolution therefore functions as a learning process in which populations acquire information through differential survival and reproduction \cite{frank2009}. Natural selection extends Schr\"odinger's picture of hereditary information by explaining how that information can accumulate and adapt over time.

Natural selection operates in systems maintained far from thermodynamic equilibrium. Replication requires the consumption of free energy and the dissipation of heat, while selection determines which information-bearing structures persist and propagate. From this perspective, evolution couples energy flow, information storage, and replication into a single nonequilibrium process that generates adaptive biological organization \cite{goldenfeld2011,england2013}.

\section{Six Principles of Adaptable Life}
Life can be understood as a consequence of sustained entropy flux across scales, from cosmological initial conditions to planetary energy dissipation \cite{schrodinger1944,kleidon2010basic,ferreira2017live}. 
The low-entropy radiation emitted by the Sun is absorbed and reprocessed by the Earth into a higher-entropy spectrum, establishing a continuous thermodynamic gradient that drives nonequilibrium processes \cite{kleidon2010basic}. 
A fraction of this flux is captured by photosynthetic systems and converted into chemical free energy, stored in molecular species maintained far from equilibrium \cite{fleming2012design}. 
These include phosphorylated nucleotides such as ATP/ADP and GTP/GDP, whose concentration imbalances sustain persistent fluxes within cellular networks \cite{qian2007phosphorylation,alberts2015molecular}. 
Biological activity thus emerges as a structured set of counter-flows embedded within a global entropy-producing process \cite{engel2007evidence}.

Within this framework, evolution selects for processes that can persist under fluctuating environmental conditions \cite{smith1982evolution,england2015dissipative}. 
Persistence requires not only stability but adaptability: systems must encode, store, and act on information about their surroundings \cite{hopfield1994physics,parrondo2015thermodynamics,wolpert2019stochastic}. 
At the molecular level, this functionality is naturally associated with switches. These are discrete elements capable of transitioning between states in response to inputs \cite{jacob1961genetic,alon2006introduction,wolpert2016free}.
The central objective is therefore to identify the physical principles that constrain the architecture and dynamics of such molecular switches.

These constraints can be formulated in terms of several general requirements.

\subsection{Entropy Source}
Life is composed of patterns of information content generated by the flow of entropy; the flow of matter and energy from regions of concentration to regions of dilution \cite{jaynes1957information}.

\subsection{Longevity}
Information must persist long enough to be functionally relevant. 
In stochastic systems, this implies the existence of metastable states separated by energy barriers that suppress rapid equilibration \cite{hanggi1990reaction,seifert2012stochastic}. 
Genetic polymers such as DNA and RNA exemplify this principle, storing information over timescales that far exceed those of individual cellular processes \cite{schrodinger1944,alberts2015molecular}. 
However, their intrinsic stability limits their responsiveness, rendering them unsuitable as fast-acting switches \cite{alon2006introduction,ptashne2011principles}.

\subsection{Fast response}
Biological systems must react to environmental changes on timescales of seconds to minutes \cite{alon2006introduction}. 
This excludes mechanisms based solely on transcriptional regulation and instead points to protein-level processes, where conformational changes and binding interactions enable rapid, reversible switching \cite{kirschner1998evolvability,pawson2000protein}. 
Such systems are embedded within signaling networks that propagate information efficiently across the cell.

\subsection{Repeatability}
Functional switches must operate cyclically. 
After transmitting information, they must reset to an initial state, enabling repeated use \cite{monod1971chance}. 
This requirement places them within the domain of nonequilibrium thermodynamics: switches function as driven cycles that consume chemical work and dissipate heat \cite{hill1989free,qian2005cycle,seifert2012stochastic,wolpert2019stochastic}. 
Sustained operation therefore necessitates the breaking of detailed balance and the maintenance of nonzero thermodynamic affinities, ensuring directed probability flux through state space \cite{polettini2014irreversible}.

\subsection{Efficiency}
Evolutionary selection imposes constraints on energetic performance. 
Switches must reliably process information while minimizing unnecessary dissipation \cite{barlow1961possible,tkacik2008information,still2012thermodynamics,mehta2012energetic,wolpert2020thermodynamic}. 
Excess energy consumption without functional gain is disfavored, implying that biological switches operate near regimes that balance speed, accuracy, and energetic cost \cite{hill2013free}.

\subsection{Multiple Switches}
The processing of more than a single bit of information requires networks of interacting switches \cite{hopfield1982neural,tkacik2008information}. 
Such networks exhibit collective behavior analogous to coupled bistable systems, where information is stored and manipulated in distributed states rather than isolated elements.

\section{Implications and Observations of the Principles}

The entropy-source principle implies that biological organization should arise wherever persistent gradients of matter and energy are available to drive nonequilibrium processes. This prediction is observed in photosynthesis, where organisms harvest solar energy to sustain metabolic activity, and in chemosynthetic ecosystems that exploit geochemical disequilibria in the absence of sunlight \cite{schrodinger1944,goldenfeld2011}.

The principle further predicts that biological order should be maintained by coupling information processing to entropy-producing flows. A canonical example is the proton gradient across cellular membranes, whose dissipation through ATP synthase powers much of cellular metabolism \cite{nelson2021lehninger}. Likewise, information-processing processes such as DNA replication, transcription, and signal transduction require continual free-energy consumption to maintain reliable states in the presence of thermal fluctuations \cite{landauer1961irreversibility,parrondo2015thermodynamics}.

Finally, the principle suggests that evolution should favor mechanisms that exploit increasingly powerful entropy sources. One of the most important examples is the emergence of oxygenic photosynthesis, which dramatically increased the free energy available to the biosphere and enabled the evolution of more complex forms of biological organization \cite{smith2016major,goldenfeld2011}.

The longevity principle implies that biological information should be stored in metastable states whose lifetimes match the timescales over which the information remains useful. This hierarchy is observed throughout biology. DNA provides evolutionary memory that persists across generations \cite{schrodinger1944,alberts2015molecular}, epigenetic modifications provide cellular memory that maintains cell identity through repeated divisions \cite{allis2015epigenetics,bird2007perceptions}, signaling states such as protein phosphorylation and allosteric conformations provide physiological memory over seconds to hours \cite{ptashne2011principles,changeux2012allostery}, and persistent neural activity and synaptic plasticity provide cognitive memory over timescales ranging from seconds to decades \cite{kandel2013principles,abbott2000synaptic}. Together, these examples suggest that biological information is universally embodied in metastable states whose persistence is tuned to functional relevance.

The fast-response principle implies that biological systems should employ information-processing mechanisms whose response times are much shorter than those of gene expression. This hierarchy is observed throughout biology. Protein conformational changes and ligand-binding events provide molecular responses on microsecond to millisecond timescales \cite{changeux2012allostery,hille2001ion}, post-translational modifications such as phosphorylation provide cellular responses on second to minute timescales \cite{pawson2000protein,hunter2000signaling}, and signaling networks built from these components enable coordinated physiological responses across entire cells and tissues \cite{alon2006introduction}. In nervous systems, ion channels and synaptic transmission support behavioral responses on millisecond timescales, allowing organisms to react rapidly to environmental stimuli \cite{kandel2013principles,hille2001ion}. Together, these examples suggest that biological regulation relies on a hierarchy of fast, reversible switches whose response times are tuned to the speed of the environmental challenges they address.

The repeatability principle implies that biological switches should operate as driven cycles that repeatedly consume free energy, perform a function, and reset. This pattern is observed throughout biology. At the molecular scale, ATPases and molecular motors such as myosin, kinesin, and dynein execute cyclic sequences of binding, conformational change, and product release that convert chemical work into directed motion \cite{alberts2015molecular,howard2001mechanics}. At the cellular scale, GTPases, phosphorylation cycles, and ion pumps function as reusable signaling and transport switches that return to their initial states after each cycle \cite{qian2005cycle,ptashne2011principles}.

The same principle appears at larger scales. The cell cycle repeatedly orchestrates DNA replication and division \cite{morgan2007cell}, circadian clocks generate self-sustained oscillations through driven biochemical cycles \cite{goldbeter1995model}, and neuronal action potentials arise from cyclic opening and closing of ion channels powered indirectly by ATP-dependent ion gradients \cite{hille2001ion}. Together, these examples suggest that biological function is organized around nonequilibrium cycles whose continual operation requires free-energy consumption, dissipation, and the maintenance of directed probability fluxes away from thermodynamic equilibrium \cite{hill1989free,seifert2012stochastic,polettini2014irreversible}.

The efficiency principle implies that biological information-processing systems should operate near tradeoffs between speed, accuracy, and energetic cost. This pattern is observed throughout biology. At the molecular scale, enzymes and molecular motors achieve high specificity and reliability while consuming only modest amounts of chemical free energy per operation \cite{alberts2015molecular,howard2001mechanics}. At the cellular scale, sensory and signaling networks often transmit information with surprisingly small numbers of molecules, suggesting selection for economical use of metabolic resources while maintaining functional accuracy \cite{tkacik2008information,alon2006introduction}.

The same principle appears at larger scales. Neural systems are among the most metabolically expensive tissues in the body, yet neural coding strategies frequently maximize information transmission per unit energy consumed \cite{barlow1961possible,laughlin2001energy}. Similarly, evolutionary innovations such as oxidative phosphorylation dramatically increased the amount of useful work obtained from available resources, enabling more complex forms of biological organization without proportionally increasing energetic expenditure \cite{lane2015vital}. Together, these examples suggest that biological systems are organized around energetic tradeoffs that favor efficient information processing rather than the maximization of any single performance metric \cite{still2012thermodynamics,mehta2012energetic,wolpert2020thermodynamic}.

The multiple-switches principle implies that biological information processing should be organized through networks of interacting switches whose collective behavior exceeds the capabilities of any individual element. This pattern is observed throughout biology. At the molecular scale, transcription factors, signaling proteins, and allosteric regulators form gene-regulatory and signaling networks that integrate multiple inputs and generate coordinated cellular responses \cite{alon2006introduction,ptashne2011principles}. At the cellular scale, interacting pathways such as MAPK, calcium, and GPCR signaling networks process information through distributed states rather than isolated molecular events \cite{pawson2000protein,kolch2005meaning}.

The same principle appears at larger scales. Neural computation emerges from networks of interconnected neurons whose collective activity stores and processes information in distributed patterns \cite{hopfield1982neural,kandel2013principles}. Similarly, multicellular organisms and ecosystems exhibit coordinated behaviors that arise from interactions among many communicating components rather than centralized control \cite{camazine2003self,levin1998ecosystems}. Together, these examples suggest that biological information is generally encoded, processed, and stored through collective dynamics of interacting switches operating across multiple scales of organization \cite{hopfield1982neural,tkacik2008information}.

\section{Discussion}

The central objective of this work has been to identify a minimal set of physical principles sufficient to support adaptable life. Building on Schr\"odinger's insight that living systems maintain ordered states through nonequilibrium energy flows, and on subsequent interpretations of evolution as an information-acquisition process, we proposed six principles governing the operation of adaptive biological systems: an entropy source, longevity, fast response, repeatability, efficiency, and multiple interacting switches. These principles were introduced as postulates rather than derived consequences and were evaluated by examining whether their implications are reflected in observed biology.

A striking feature of the resulting framework is the recurrence of the same organizational motifs across biological scales. Long-lived information appears as genetic, epigenetic, physiological, and cognitive memory. Fast responses emerge through protein conformational changes, signaling cascades, and neural activity. Repeatable operation is achieved through molecular, cellular, and organismal cycles. Information processing arises not from isolated switches but from networks of interacting elements whose collective behavior exceeds that of any individual component. The repeated appearance of these motifs suggests that they are not accidental features of particular biological systems but generic consequences of the physical requirements imposed by adaptation.

The framework also highlights the intimate relationship between information and thermodynamics. Information storage requires metastability. Information transmission requires rapid state transitions. Information processing requires networks of interacting switches. All of these functions depend on continual free-energy consumption and entropy production. From this perspective, biological information is not separate from thermodynamics but is a manifestation of nonequilibrium thermodynamic organization. Adaptation therefore emerges from the interplay of energy flow, information storage, and natural selection, rather than from any one of these factors in isolation.

An important limitation of the present work is that the six principles remain qualitative. They organize a wide range of biological observations, but they do not yet constitute a predictive theory. Future work should seek quantitative formulations of each principle and determine whether they impose measurable constraints on biological systems. Examples include limits on information lifetime, tradeoffs between switching speed and energetic cost, bounds on network complexity, and relationships between thermodynamic dissipation and adaptive performance. Such developments would allow the framework to move beyond classification toward prediction.

The broader significance of the approach lies in its attempt to extend Schr\"odinger's program into the era of nonequilibrium statistical physics and information theory. Schr\"odinger explained how hereditary information can persist in matter. Modern evolutionary theory explains how information accumulates through natural selection. The present work suggests that adaptability itself may be understood as the consequence of a small set of physical constraints governing the storage, transmission, and processing of information in systems driven far from thermodynamic equilibrium. Whether these six principles are sufficient remains an open question, but they provide a starting point for a physics-based theory of adaptable life.

%

%
\bibliographystyle{splncs04}
\bibliography{library}
%

%
%

\end{document}